\input harvmac

\def\IB{\relax\hbox{$\inbar\kern-.3em{\rm B}$}}
\def\IC{\relax\hbox{$\inbar\kern-.3em{\rm C}$}}
\def\ID{\relax\hbox{$\inbar\kern-.3em{\rm D}$}}
\def\IE{\relax\hbox{$\inbar\kern-.3em{\rm E}$}}
\def\IF{\relax\hbox{$\inbar\kern-.3em{\rm F}$}}
\def\IG{\relax\hbox{$\inbar\kern-.3em{\rm G}$}}
\def\IGa{\relax\hbox{${\rm I}\kern-.18em\Gamma$}}
\def\IH{\relax{\rm I\kern-.18em H}}
\def\IK{\relax{\rm I\kern-.18em K}}
\def\IL{\relax{\rm I\kern-.18em L}}
\def\IP{\relax{\rm I\kern-.18em P}}
\def\IR{\relax{\rm I\kern-.18em R}}
\def\IZ{\relax\ifmmode\mathchoice
{\hbox{\cmss Z\kern-.4em Z}}{\hbox{\cmss Z\kern-.4em Z}}
{\lower.9pt\hbox{\cmsss Z\kern-.4em Z}}
{\lower1.2pt\hbox{\cmsss Z\kern-.4em Z}}\else{\cmss Z\kern-.4em Z}\fi}






\def\unlockat{\catcode`\@=11}
\def\lockat{\catcode`\@=12}

\unlockat

\def\newsec#1{\global\advance\secno by1\message{(\the\secno. #1)}
\global\subsecno=0\global\subsubsecno=0\eqnres@t\noindent
{\bf\the\secno. #1}
\writetoca{{\secsym} {#1}}\par\nobreak\medskip\nobreak}
\global\newcount\subsecno \global\subsecno=0
\def\subsec#1{\global\advance\subsecno
by1\message{(\secsym\the\subsecno. #1)}
\ifnum\lastpenalty>9000\else\bigbreak\fi\global\subsubsecno=0
\noindent{\it\secsym\the\subsecno. #1}
\writetoca{\string\quad {\secsym\the\subsecno.} {#1}}
\par\nobreak\medskip\nobreak}
\global\newcount\subsubsecno \global\subsubsecno=0
\def\subsubsec#1{\global\advance\subsubsecno by1
\message{(\secsym\the\subsecno.\the\subsubsecno. #1)}
\ifnum\lastpenalty>9000\else\bigbreak\fi
\noindent\quad{\secsym\the\subsecno.\the\subsubsecno.}{#1}
\writetoca{\string\qquad{\secsym\the\subsecno.\the\subsubsecno.}{#1}}
\par\nobreak\medskip\nobreak}

\def\subsubseclab#1{\DefWarn#1\xdef
#1{\noexpand\hyperref{}{subsubsection}%
{\secsym\the\subsecno.\the\subsubsecno}%
{\secsym\the\subsecno.\the\subsubsecno}}%
\writedef{#1\leftbracket#1}\wrlabeL{#1=#1}}
\lockat

\def\IL{\relax{\rm I\kern-.18em L}}
\def\IH{\relax{\rm I\kern-.18em H}}
\def\IR{\relax{\rm I\kern-.18em R}}

\def\IZ{\relax\ifmmode\mathchoice
{\hbox{\cmss Z\kern-.4em Z}}{\hbox{\cmss Z\kern-.4em Z}}
{\lower.9pt\hbox{\cmsss Z\kern-.4em Z}}
{\lower1.2pt\hbox{\cmsss Z\kern-.4em Z}}\else{\cmss Z\kern-.4em
Z}\fi}


\def\IZ{\relax\ifmmode\mathchoice
{\hbox{\cmss Z\kern-.4em Z}}{\hbox{\cmss Z\kern-.4em Z}}
{\lower.9pt\hbox{\cmsss Z\kern-.4em Z}}
{\lower1.2pt\hbox{\cmsss Z\kern-.4em Z}}\else{\cmss Z\kern-.4em
Z}\fi}
\def\IB{\relax{\rm I\kern-.18em B}}
\def\IC{{\relax\hbox{$\inbar\kern-.3em{\rm C}$}}}
\def\ID{\relax{\rm I\kern-.18em D}}
\def\IE{\relax{\rm I\kern-.18em E}}
\def\IF{\relax{\rm I\kern-.18em F}}
\def\IG{\relax\hbox{$\inbar\kern-.3em{\rm G}$}}
\def\IGa{\relax\hbox{${\rm I}\kern-.18em\Gamma$}}
\def\IH{\relax{\rm I\kern-.18em H}}
\def\II{\relax{\rm I\kern-.18em I}}
\def\IK{\relax{\rm I\kern-.18em K}}
\def\IP{\relax{\rm I\kern-.18em P}}

\def\inbar{\,\vrule height1.5ex width.4pt depth0pt}

\font\cmss=cmss10 \font\cmsss=cmss10 at 7pt
\def\IR{\relax{\rm I\kern-.18em R}}


\def\boxit#1{\vbox{\hrule\hbox{\vrule\kern8pt
\vbox{\hbox{\kern8pt}\hbox{\vbox{#1}}\hbox{\kern8pt}}
\kern8pt\vrule}\hrule}}
\def\mathboxit#1{\vbox{\hrule\hbox{\vrule\kern8pt\vbox{\kern8pt
\hbox{$\displaystyle #1$}\kern8pt}\kern8pt\vrule}\hrule}}


\def\inbar{\,\vrule height1.5ex width.4pt depth0pt}

\font\cmss=cmss10 \font\cmsss=cmss10 at 7pt
\def\IR{\relax{\rm I\kern-.18em R}}

%
\def\hh{hep-th/}
\lref\horava { P. Horava,  { \it M-Theory as a Holographic Field Theory}, hep-th/9712130.}

\lref\holography { G. 't Hooft, {\it Dimensional Reduction in Quantum Gravity }
in {\it Salamfest} 1993, p.284, gr-qc/9310026; L. Susskind, {\it The world as a Hologram }, J. Math. Phys.36 (1995).}

\lref\sugragauging{ { L. Castellani, R. D'Auria, P. Fre,
\it Supergravity and Superstrings, World Scientific 1991}; M. Duff and
B. Nilsson {\it  Kaluza-Klein Supergravity}, 
Phys. Rep. 130 (1986) 1.}

\lref\seiberglist{N. Seiberg, {\it Five Dimensional SUSY Field Theories, Non-trivial Fixed
Points and String Dynamics}, hep-th/9608111, Phys.Lett. B388 (1996) 753}.

\lref\natifix {A partial list of recent works is:  E.Witten, {\it Some Comments on String Dynamics},
Contribution to STRINGS95: Future Prospects in String Theory, Los Angeles, CA,
13-18 March 1995, hep-th/9507121 \semi
O. Ganor and A. Hanany, {\it Small E(8) 
Instantons and Tensionless Noncritical strings}, hep-th/9602120,
Nucl.Phys. B474 (1996) 122\semi  N. Seiberg and E. Witten,{ \it Comments on String
Dynamics in Six Dimensions}, hep-th/9603003, Nucl.Phys. B471 (1996) 121\semi 
N. Seiberg, {\it Five Dimensional SUSY Field Theories, Non-trivial Fixed
Points and String Dynamics}, hep-th/9608111, Phys.Lett. B388 (1996) 753\semi
N. Seiberg, {\it Non-trivial Fixed Points of The Renormalization Group in
Six Dimensions}, hep-th/9609161, Phys.Lett. B390 (1997) 169\semi
M. Bershadsky and C. Vafa, {\it Global Anomalies and Geometric Engineering
of Critical Theories in Six Dimensions}, hep-th/9703167\semi 
J.L. Blum and K. Intriligator, {\it New Phases of String Theory and 6d RG
 Fixed Points via Branes at Orbifold Singularities}, hep-th/9705044, 
Nucl.Phys. B506 (1997) 199\semi
N. Seiberg, {\it Notes on Theories with 16 Supercharges}, hep-th/9705117, 
An updated version of lectures presented at the Jerusalem Winter School on
Strings and Duality. To appear in the Proceedings of the Trieste Spring School
(March 7--12, 1997).}

\lref\simons{  J. Cheeger and J. Simons, {\it Differential Characters and Geometric Invariants},
 Stony Brook Preprint, (1973), unpublished.}

\lref\townsend{  E. Bergshoeff, L.A.J London and
     P.K. Townsend, {\it Spacetime Scale-Invariance and the Super p-Brane},
 hep-th/9206026, Class. Quant. Grav. 9 (1992) 2545.}

 \lref\seibergsix{  O. Aharony, M. Berkooz, N. Seiberg,  {\it Light-Cone Description of (2,0) Superconformal Theories in Six Dimensions},
  hep-th/9712117\semi  O. J. Ganor, David R. Morrison, N. Seiberg,
 {\it
Branes, Calabi-Yau Spaces, and Toroidal Compactification of the N=1 Six-Dimensional $E_8$ Theory}, hep-th /9610251, Nucl. Phys.  B487 (1997) 93-127\semi 
N.  Seiberg, 
{\it Non-trivial Fixed Points of The Renormalization Group in Six Dimensions},  hep-th/9609161, Phys. Lett.  B390 (1997) 169-171. 
}
    
 \lref\wittensix{E.  Witten, {\it New  Gauge  Theories In Six Dimensions},
  hep-th/9710065. }

\lref\laroche{  L.~Baulieu, C.~Laroche, { \it 
On Generalized Self-Duality Equations Towards Supersymmetric Quantum Field Theories Of Forms
},  hep-th/9801014.}

\lref\west{  L.~Baulieu, P.~West,
{  \it Six Dimensional TQFTs and  Self-dual Two-Forms.
 }}

\lref\bks{  L.~Baulieu, H.~Kanno, I.~Singer, 
{\it Special Quantum Field Theories in Eight and Other Dimensions}, hep-th/9704167, Talk given at
APCTP Winter School on Dualities in String Theory, (Sokcho, Korea),
February 24-28, 1997
\semi  
  L.~Baulieu,  H.~Kanno, I.~Singer, {\it Cohomological Yang--Mills
Theory
in Eight Dimensions}, hep-th/9705127, to appear in Commun. Math. Phys.}

\lref\wittentopo { E. Witten,  {\it  Topological Quantum Field Theory},
Commun. Math. Phys. {117}(1988)353. }

\lref\wittentwist { E. Witten, {\it Supersymmetric Yang--Mills theory on a
four-manifold},  \hh9403195, J. Math. Phys. {35} (1994) 5101.}

\lref\sugragauging{ see for example{ L. Castellani, R. D'Auria, P. Fre,
\it Supergravity and Superstrings, World Scientific 1991}\semi M. Duff and
B. Nilsson, {\it  Kaluza-Klein Supergravity}, 
Phys. Rep. 130 (1986) 1.}

\lref\bs { L. Baulieu, I. M. Singer, {\it Topological Yang--Mills
Symmetry}, Nucl. Phys. Proc. Suppl.  
15B (1988) 12\semi  L. Baulieu, {\it On the Symmetries of Topological Quantum
Field Theories},   hep-th/
9504015, Int. J. Mod. Phys. A10 (1995) 4483\semi R. Dijkgraaf, G. Moore,
 {\it Balanced Topological Field Theories},
hep-th/9608169,   Commun. Math. Phys. 185 (1997) 411.}

\lref\kyoto { L. Baulieu, {\it Field Antifield Duality, p-Form Gauge Fields and Topological Quantum
Field Theories}     hep-th/9512026,   Nucl. Phys. B478 (1996) 431. } 
   
 \lref\strings {  L. Baulieu, M. B. Green, E. Rabinovici {\it A Unifying
Topological Action for Heterotic and  Type II Superstring  Theories},
hep-th/9606080, Phys.Lett. B386 (1996) 91;
{\it   Superstrings from   Theories with $N>1$ World Sheet Supersymmetry},
 hep-th/9611136, Nucl. Phys. B498 (1997). }

\lref\sourlas{  G. Parisi and N. Sourlas,
{\it Random Magnetic Fields, Supersymmetry and Negative Dimensions},  Phys.
Rev.Lett. 43 (1979) 744;  Nucl. Phys. B206 (1982) 321. }

\lref\SalamSezgin{A. Salam  and  E. Sezgin,
{\it Supergravities in diverse dimensions}, vol. 1, p.119;
P. Howe, G. Sierra and P. Townsend, Nucl Phys B221 (1983) 331}.

\lref\nekrasov{ A. Losev, G. Moore, N. Nekrasov, S. Shatashvili,
{\it
Four-Dimensional Avatars of Two-Dimensional RCFT},  hep-th/9509151,
Nucl. Phys. Proc. Suppl.  46 (1996) 130; L. Baulieu, A. Losev, N.
Nekrasov  {\it Chern-Simons and Twisted Supersymmetry in Higher
Dimensions},  hep-th/9707174, to appear in Nucl. Phys. B.  }

 \Title{ \vbox{\baselineskip12pt\hbox{hep-th/9805122}
\hbox{CERN-TH: 98-96}
\hbox{RI: 98-10}
\hbox{LPTHE: 98-17  }}}
{\vbox{
 \centerline{ Self-Duality and   New TQFTs  for Forms  }}}
\centerline{Laurent Baulieu   }
\vskip 0.08cm
\centerline{CERN, 
Geneva, Switzerland}
\vskip 0.08cm
\centerline{and}
\centerline{LPTHE\foot{UMR CNRS associ\'ee aux  Universit\'es Pierre et Marie Curie (Paris VI) et Denis Diderot (Paris~VII).}, Paris, France.}

\vskip 0.15cm

 \medskip
\centerline{ Eliezer Rabinovici \foot{Work supported in part by the Israel Academy of Sciences and Humanities,  Center of Excellence Program and the American--Israeli Binational Science Foundation-BSF.}   }
\vskip 0.1cm
\centerline{CERN, 
Geneva, Switzerland }
\vskip 0.1cm
\centerline{and}
\centerline{  Racah Institute of Physics,  Hebrew University, Jerusalem,
Israel. }

 \vskip 0.5cm

We discuss theories containing higher-order forms in various dimensions. 
We  explain how Chern--Simons-type theories of 
forms can be defined from  TQFTs in one less dimension.  We  also  exhibit 
 new TQFTs with
interacting Yang--Mills fields and higher--order forms. They are 
 obtained   by the dimensional 
reduction of  TQFTs whose gauge functions are   free self-duality 
equations. Interactions are due
to  the  gauging of  global internal symmetries after  dimensional reduction. We list   possible
symmetries and give a brief discussion on the possible relation of such 
systems to interacting field theories.


\def\e{\epsilon}
\def\demi{{1\over 2}}
\def\pa{\partial}
\def\a{\alpha}

\def\m{\mu}
\def\n{\nu}

\def\X{X}

\def\H{H}

\def\do{{D+1}}
 \Date{ }

 \def\I{{\it I}}
 

 \newsec{Introduction}

Recent interesting results shed   new light on
field theories in dimensions higher than four which contain no explicit
graviton interactions and involve higher--order forms. Motivated by a new understanding of non perturbative
features of string theory and insights obtained for an M theory, 
a rich structure of   supersymmetric theories was uncovered in five and
six dimensions   \natifix. Another set of results concerns the exploration of the holography principle. According to that principle the degrees of freedom of some higher-dimensional systems are actually fully accounted for by degrees of freedom residing only  on the boundary of those systems \holography. 
It has indeed been suggested \horava\ that   $p$-form gauge fields  with  $p>1$  could be used to  implement   these ideas.

In this paper we will study theories of   forms   in an attempt to enrich the class of models depending on such fields.
Our starting point is Topological Field Theory (TQFT). TQFTs are  quantum field theories that  possess observables depending  only  on  global degrees of
freedom.    Examples are the
cases of the four and eight dimensional topological Yang--Mills theory
\wittentopo\bs\bks. In these cases, the  topological behaviour is particularly clear because the TQFT actions can be described as BRST-invariant gauge-fixings of classical topological invariants. Moreover, they have the interesting feature that they are   linked by twist to supersymmetric theories with physical
particles.
For  those of the  higher-dimensional TQFTs which can be untwisted into theories with Poincar\'e supersymmetry,   one could hope that,  the latter could inherit from the former  properties which allow them to  pass the barrier of naive non renormalizability.

It must be noted that the  TQFTs
depending on forms with $p>1$ are likely to have a rich content because, at least in the abelian case, it is known that a $p$-form can be defined
with a status analogous to that of a connection, so that one expects non vanishing topological invariants when one integrates exterior products of their curvatures over certain spaces (see e.g. \simons).

This work has two aspects.  The first part   aims at finding new
Chern-Simons (CS) like theories, and defining them. What we mean by a {
{classical}} CS-type action is an action made of exterior products of  $
p$-forms  and their curvatures   (the  Yang--Mills Chern--Simons actions are particular cases).   It is independent of the
metric  and of first order and may have     gauge invariances for forms. The    Hamiltonian vanishes modulo gauge
transformations. It was   suggested that   CS theories   embody the
holography principle.    We     show that   the procedure initiated in
\nekrasov\    gives a   method for  defining such theories.   Given a  TQFT in $D$ dimensions, based on the BRST quantization of a topological
term, we observe  that it   can   often be put in correspondence with  a CS-type
theory   in $ (D+1)$ dimensions. The way we define the CS theory implies that it is    invariant  under  a symmetry  operation $Q$, such that
 \eqn\auto{\eqalign{Q \varphi=\Psi}}   
and that there is a mapping between the fields of the TQFT in $D$ dimensions and
those of the CS theory in $ (D+1)$ dimensions.

The CS action decomposes into a $Q$-invariant but not $Q$-exact part and a 
$Q$-exact part. The $Q$-exact part can be seen as a topological gauge-fixing
term, analogous to that occurring in the $D$-dimensional theory. Moreover, the  
topological  gauge functions  in $(D+1)$ dimensions   are inspired by those of
the  TQFT in $D$ dimensions\foot{ The topological  gauge functions are gauge-invariant under the
ordinary gauge symmetries of forms: they determine a gauge invariant Lagrangian
whose residual ordinary gauge-invariance must  be gauge-fixed, eventually by the
standard methods.}. Their role is  to    provide    in a $Q$-exact  way terms such that
the complete action is  second order   for the bosons  and first order for the
fermions.  The $Q$-invariant but not $Q$-exact part contains a classical CS
term, plus supersymmetric terms. It  exists because  the
cohomology with ghost number zero of $Q$  is   not empty  in     $ (D+1)$ dimensions, contrary to what occurs  in    $D$ dimensions.   

  The
symmetry \auto\  indicates a gauge invariance of the topological type. Its existence simplifies the counting of local degrees of freedom of the CS action: by establishing a  pairing  between  the  bosons $\varphi$ and fermions $\Psi$, it    allows an  investigation of 
the topological properties of the theory, at least formally.
As a matter of fact,  the supersymmetric formalism for the CS-type theory in $(D+1)$ dimensions establishes almost by definition that the
theory   has no   gauge-invariant local degrees of freedom in the "bulk", since only
global excitations can remain in the Hilbert space defined by the principle
of $Q$-invariance. This agrees with the property that the classical
Hamiltonian vanishes   (modulo gauge transformations).
  In the absence of such a symmetry, the counting of degrees of freedom for  
  TQFTs based on CS   first-order  actions is more involved, except in the three -dimensional Chern--Simons Yang--Mills case where one can directly   prove   that  the theory does indeed  describe only 
global degrees of freedom  (in particular  by  solving Gauss's law).
Other gauges exist, which can be enforced in a $Q$-invariant way, such that   all fermions decouple, and the action formally reduces  to a classical bosonic   
CS action, whose direct quantization is challenging. 

We can tentatively give a physical interpretation to this quite general  
mapping between     $D$- and  $(D+1)$-dimensional models. What we obtain is an
identification between   the off-shell  degrees of freedom of two theories, in
$D$ and   $(D+1)$ dimensions. Their  Lagrangians are  usually not related  by 
ordinary dimensional reduction. However, their $Q$-exact parts are related by
the choice of the topological gauge functions.  Keeping in mind that the
D-dimensional theory is a TQFT that  can often  be untwisted into   a physical 
supersymmetric theory    (e.g. a theory of particles with  the physical spin
statistics relation
 identified as the  gauge-invariant untwisted fields), we might  not be very far from
the idea of the holography principle: there is  a physical theory with propagating
particles    on   the   boundary of the  $(D+1)$-dimensional space, in which the
theory      takes the form  of a topological CS-type,one with no
particles.

In a second part, we   construct classes of theories of higher--dimensional forms
for which one can apply the general formalism relating  TQFTs in $D$ and ($D$+1)
dimensions:  these yet unexplored TQFTs are obtained  by  combining
new types of
free self-duality equations with ordinary dimensional reduction. They    generalize  the simplest examples 
$dB_k=*dB_k$ for a single field in $D=2k+2$ dimensions  into $dB _p=^* dC
_{D-2-p}$, where both fields 
$B$ and $C$ are independent, with no restriction on $p$, except,
$p<D-1$; by dimensional reduction in a lower dimension  $d$,  
$B_p$ and $C _{D-2-p}$ generate   a tower of forms $B^i_q$ and 
$C^i_{d-2-q}$, with $q\leq p$. One has, as a starting point, an  invariant
of the type $\int _D dB_p\wedge dC _{D-2-p}$. Because of \simons, 
we expect that this topological invariant be non zero (it is ${ \it Z}$-valued, up to an overall normalization).  The free self-duality equations are   
$dB^i_q=^* dC^i_{d-2-q}$. 
The  forms $B^i_q$ and 
$C^i_{d-2-q}$ can be arranged into representations of
certain Lie algebras, so that  
$i$ can be seen as  a global internal symmetry    index. We  then observe
that a gauging of the global symmetries can  often be done, and this allows us
 to
modify the  self-duality equations. We eventually      get   TQFTs involving   
Yang--Mills fields and forms, which interact by minimal coupling through the "gauged"
self-duality equations: 
\eqn\tarata{\eqalign{DB^i_q=^* DC^i_{d-2-q}
\quad\quad\quad
\eqalign{F_2^a =^* DC^a_{d-3}} }}
 Here $D$ is the covariant derivative with respect to the  Yang--Mills field made from the one-forms    obtained by the process of dimensional reduction, while $F_2 $ is its curvature. In four dimensions, the last equation is $F_2=^*F_2$.

In \west, it was shown that the notion of twist of
topological symmetries goes beyond the case of the four-dimensional Yang--Mills
field theory, for instance in the case of TQFTs with two-forms in six
dimensions, and three-forms in eight
dimensions, with self-duality equations. It is thus an appealing possibility that the TQFTs with gauged
internal supersymmetries   exhibited in this paper can be untwisted into  new types of
theories, with reduced Poincar\'e supersymmetry. The   number of supersymmetry generators
  might be  less than in the standard theories, because of      
interactions. In an extremal situation,   the untwisted supersymmetry 
has  only one   
generator (corresponding to the BRST
$Q$-operator). On the other hand, the cases of the  four- and eight-dimensional Yang--Mills TQFTs show that  the full Poincar\'e 
supersymmetry can be  obtained  after untwisting.  We list possible internal gauge
symmetries for the TQFTs in four and five dimensions, obtained by dimensional
reductions of TQFTs in dimensions up to 14.
Such a list might be   useful for comparison with the list of possible global symmetries of
non-trivial interacting field theories in five dimensions \seiberglist.
 Independently of   the possible  interpretations of
these  internal symmetries, the self-duality
equations that we have found  indicate the existence of  yet unexplored
   TQFTs.
  They are very
natural generalizations of the four-dimensional Yang--Mills TQFT.  Finally, the relation of
such  TQFTs with theories with Poincar\'e  
  supersymmetry is a very interesting question, which  stimulates
us to     understand    the notion of twist in quantum field
theory dynamically.

\newsec{Transgression between D- and (D+1)-dimensional TQFTs}

 The aim of this section is     to indicate   an     algorithm for defining a 
quantum field theory  related to
first-order actions  in $(D+1)$ dimensions.  One  starts from 
an "ordinary" TQFT
in
$D$ dimensions. It is defined from the knowledge of a BRST operator
$Q$ acting on a set of fields, for instance $p$-form gauge fields, their 
 ghosts and
topological ghosts. $Q$ and the $Q$-invariant     action  $\I_D$ 
satisfy   
\eqn\nild{Q^2=0 \  {\rm modulo \ gauge \ transformations \ of \ forms\ in} \ D \ {\rm  dimensions}
}
 \eqn\actiond{
\I_D=  \I_{D   \ {\rm top }}[\varphi] + \big\{Q, \ldots\}
}
 where $ \I_{D  \  {\rm{top }}}$ is a classical topological invariant, which is a function of only the classical fields $\varphi$, so the
classical Lagrangian is  locally   a  $d$-exact $D$-form. 
The set of fields on
which $Q$ acts can be found from methods as in \bks\west, and the terms denoted as
$\ldots$ in the $Q$-exact term are topological gauge functions as those
 defined,   e.g
  in \laroche.

Our observation is that the fields of the $D$-dimensional TQFT can be redefined 
as elements of a multiplet in ($D$+1)-dimensional space, with a
modified definition of $Q$:   $Q$ acts on  the classical fields of the ($D$+1)-dimensional theory  as in \auto\ and  it        satisfies on all
fields  the following relation:   
 \eqn\nildd{Q^2=\pa_\do \  {\rm modulo\ gauge\ transformations\
of \ forms \ in}\ (D+1)\ {\rm dimensions.}} 
Moreover, one has   a $Q$-invariant
action\eqn\actiondd{I_{ D+1 }=  \I_{D+1 \rm{CS
}}[\varphi,\Psi] +\big\{ Q, \ldots\}}.

The role of the    $Q$-exact term  in the right-hand side of \actiondd\ is to define  the
theory at the quantum level:  the $\ldots$ terms  are in correspondence with    the
topological gauge functions  of the TQFT  in 
  $D$ dimensions, and they eventually provide a  Lorentz-invariant contribution to  the 
action in $(D+1)$ dimensions\foot{This   holds true  despite the fact that the expression of
$Q$ violates Lorentz invariance in $(D+1)$ dimensions, as shown by \nildd.}, with a
dependence on the field derivatives   which is second order for the    bosons and first
order for the fermions.   The other term in  \actiondd, $\I_{D+1 \rm{CS }}[\varphi,\Psi]$,   
is the sum of a  classical first-order action and of a    ghost-dependent action. This part of the Lagrangian is
$Q$-invariant, but it is neither  $Q$-exact nor $d$-exact as the $D$-dimensional theory Lagrangian. The expression     
$\I_{D+1 \rm{CS }}[\varphi,0]$ can be   called the classical bosonic CS action. Without the addition of the $Q$-exact term to  $\I_{D+1 \rm{CS }}[\varphi,\Psi] $,  the
definition of the theory is  difficult, and may be ambiguous    (since the Hamiltonian of 
$\I_{D+1 \rm{CS }}[\varphi,0] $ identically vanishes).  By a suitable choice of the    $\ldots$ terms in  \actiondd,  the  $Q$-exact term  determines a meaningful local quantum theory, because it adds to $\I_{D+1 \rm{CS }}[\varphi,\Psi]$  a sum of
squared curvatures.  One can adopt the point of view that $\I_{D+1 \rm{CS
}}$
   can be factored out in
the measure of the path integral, as for instance   the $\theta$ term in $QCD$.  

One generally finds observables other than $\I_{D+1 \rm{CS
}}[\varphi,\Psi] $  in the cohomology of
$Q$ in $(D+1)$ dimensions  . They must be searched for case by case, by lifting to a   $(D+1)$-dimensional
space the $p$-forms on which $Q$ acts in $D $ dimensions.  

In what follows,  for the sake of notational simplicity, we     
  examine  the case of two-form gauge fields, and see how a CS-type theory  for two-forms
in five dimensions is related to a TQFT in four dimensions depending on two-forms {\it and
} scalars. Then, we   generalize   the construction.

Our formula holds for fields with no interactions. They can be  adapted to the cases  of systems
whose interactions are described by non-free curvatures satisfying Bianchi identities.

 \def\I{{\it I}}

\subsec{The example of two-form gauge fields}

As shown in \west, the field spectrum of a TQFT for a two-form gauge field
$B_{ij}$, where   $i,j,\ldots$ are $D$-dimensional indices is described  by
the following  TQFT multiplet
  \eqn\mul {\eqalign{(B_{ij}, \Psi_{[ij] }^{1},
\chi_{\a}^{-1},\H_{\a}  , \eta_{i }^{-1} ,\Phi^{ 3},   \Phi^{ -3},
\X^{1}, \Phi^{ 2}_i,  \Phi^{ -2}_i, \eta^{2}, \eta^{-2}) \cr }  }
The definition of    the
topological  BRST operator $Q$ is     
   \eqn\symQf{\eqalign{& Q B_{ij} =
 \Psi^1_{[ij]} \quad  Q \Psi^1_{[ij]} =\pa_{[i}\Phi^2_{j]}\cr & Q \Phi_{i }^2
= \pa_{i } \Phi^3\quad  Q \Phi^3=0 
 \cr & Q  \chi_{\a}^{-1} = \H_{\a} ^{ } \quad  Q \H_{\a} ^{ } =0\cr & Q 
\Phi_{i }^{-2} = \eta^{-1}_i  \quad  Q \eta^{-1}_i =0 
 \cr & Q  \Phi^{-3}  = \eta^{-2}\quad  Q  \eta^{-2} =0 
 \cr & Q  \X^{ 1}  = \eta^{2}\quad  Q  \eta^{2} =0 
 \cr }  } The upper index denotes the ghost number. Its  value modulo 2
determines the statistics of fields. The index $\a$ of the antighost
$\chi_{\a}^{-1}$ runs over  the number of independent topological gauge
functions  (e.g.  gauge-invariant generalized  self-duality
conditions as in \laroche)  that the TQFT imposes to define a local dynamics
for the two-form gauge field.  The TQFT action has the form\foot{We assume  the existence of     gauge functions ${\it F}_\a(B)$
without  specifying their  expressions.}:
 \eqn\actionD{\eqalign{& \I_D= \I_{\rm top}[B]
+
 \int _D \big\{Q, 
\chi_{\a}^{-1}({\it F}_\a(B)+\demi H_{\a} ) +\Phi^{-2}_i \pa_{j } 
\Psi^1_{[ij]} +\Phi^{-3}  \pa_{i }  \Phi^2_{i} +\X^{1}  \pa_{i } 
\Phi^{-2}_{i} \}
 \cr }  } The BRST algebra \symQf\ satisfies \nild.   

The action \actionD\ has been studied for  
$D=6$ in \west. In this case,   self-duality equations can be derived for a pair of two-forms and  
 thus the  gauge functions  can be determined. After the 
 elimination of fields with algebraic equations of motion, one has the interesting result
 that the TQFT symmetry can be related by a twist to     Poincar\'e supersymmetry
\west.   

The  definition of  a theory for    two-forms  
 in $(D+1)$ dimensions follows from  the possibility of  rearranging  by linear
combinations      
all fields of  the $D$-dimensional Lorentz covariant TQFT multiplet  \mul\  into a   
$(D+1)$-dimensional Lorentz-covariant multiplet  \eqn\muldd {\eqalign{(B_{\m\n},
\Psi_{[\m\n] }^{-},   \phi_{\m }^{+}, \phi ^{-},  \Phi^{ -},  
  \X^{-}, \eta^{ +} ,   \chi_{\a }^{-}, H_{\a } )
  \cr }  } 
Let us detail this multiplet. The $ (D+1) $-dimensional Greek indices 
can be   decomposed as $\do,i$. The mapping between the fields in \mul\
and \muldd\  is as follows 
\eqn\mapping {\eqalign{    B_{ij} , \Phi_{i }^{2}, \Phi_{i }^{-2},
\eta^{2}
&\to
B_{\m\n},  \phi_{\m }^{+}
  \cr
\Psi_{[ij] }^{1}, 
 \eta^{-1}_i&\to \Psi_{[\m\n] }^{-}
   \cr   \Phi ^{3},  \Phi^{ -3},  
  \X^{-1}, \eta^{ -2} ,   \chi_{\a }^{-} 
 & \to\phi ^{-}, \Phi^{ -},  
  \X^{-}, \eta^{ +} ,   \chi_{\a }^{-}
 \cr }  }
This mapping violates modulo 2 the ghost-number conservation of the $D$-dimensional
theory (this  explains our   unified notation by  an upper index $\pm$ for   the positive
and negative statistics of fields in \muldd).  The   BRST operator $Q$, which satisfies  
\nildd\ , is
 \eqn\symQdd{\eqalign{ Q B_{\m\n}& =
 \Psi^-_{[\m\n]}
\quad   \quad  \quad\quad\quad\quad\quad
 Q \Psi^-_{[\m\n]}  = \pa_{[\do}B_{\m\n]} + \pa_{[\m}\phi^+_{\n]}
\cr   
Q \phi_{\m}^+
&= \Psi^-_{[\do\m]} + \pa_{\m} \phi^-
\quad\quad\quad 
  Q \phi^- = \phi_{\do}^+
 \cr }}
and 
 \eqn\symQddd{\eqalign{ & Q  \chi_{\a}^{-} = \H_{\a}    \quad 
 Q \H_{\a}   = \pa_\do  \chi_{\a}^{-} \cr & Q  \Phi^{-}  = \eta^{+}\quad 
Q  \eta^{+} =\pa_\do   \Phi^{-} 
 \cr & Q  \X^{ -}  = \eta^{+}\quad  Q  \eta^{+} =\pa_\do   \X^{ -}  }}

In  comparison, in the Yang--Mills case, 
 the fields of the $D$-dimensional Yang--Mills TQFT    
$(A_i,\Psi_i^1,\Phi^{2}, \Phi^{-2}, \eta^{-1},  \chi^{-1})$  can be   
rearranged into the
$(D+1)$-dimensional  multiplet $(A_\m,\Psi^-_\m,\phi^{+},  \chi^{-1})$
and the equations analogous to \symQdd\
and  \symQddd\ are \nekrasov:
\eqn\symQymdd{\eqalign{  Q A_{\m} &=
 \Psi^-_{\m} \quad \quad  
 Q \Psi^-_{\m} = F_{\do\m} + \pa_{\m}\phi^+ \cr  
 Q \phi^+ &= \Psi^-_{\do}  
 \cr   Q  \chi  ^{-} &= \H   \quad \quad   
 Q \H    = \pa_\do  \chi   \cr }}

Next comes the question of determining a $Q$-invariant action in $(D+1)$
dimensions. Equation \actiondd\ indicates   two distinguishable
terms. The $Q$-exact term exists in $(D+1)$ dimensions if    there are  generalized  self-duality equations in $D$ dimensions; the $Q$-invariant and  not $Q$-exact term
exists if there is a  Chern--Simons-like term in $(D+1)$ dimensions.
Choosing $D=4$, one finds for instance that a
minimal set of fields involving two-forms consists  of a pair of two
two-forms $B_{ij}$ and $\bar B_{ij}$  and  two zero-forms $S $ and $\bar
S$. This is    justified by the existence 
 of a  TQFT for a three-form gauge field in eight dimensions 
with ordinary self-duality   gauge conditions \bks\west\  and by 
its dimensional reduction according to \eqn\reduc{\eqalign{&
 D=8\ : \ B_3
  \cr & 
 D=7\ : \ B_3\ B_2
  \cr & 
 D=6\ : \ B_3    \ 2B_2    \ B_1
  \cr & 
 D=5\ :   \ B_3    \ 3B_2     \ 3B_1  \  S
  \cr &
 D=4\ :   \ B_3    \ 4B_2     \  6B_1  \  4S
  }}
(The  generality of this reduction will be addressed  in the last
section.)
 The set of fields for $D=4$ includes: a three-form gauge field,  which determines a TQFT by itself (it
carries no degree of freedom and can be used    to allow spontaneous symmetry breaking   in 
scale-invariant    theories of three-branes   \townsend);  six one-forms,  four two-forms and four 
zero-forms.   The dimensionally reduced self-duality equations are 
 \eqn\self{\eqalign{&
F^a_{ij}= \e _{ijkl}F  ^{a kl}
  }}
\eqn\selff{\eqalign{&
F^b_{ijk}= \e _{ijkl  }F ^{b  }_l
  }}
  where $i,j,k,l$ are now four-dimensional indices. The internal 
  indices $a $ and $b$
run over   six and four possibilities respectively. The two-forms
 $F_2^a$ are the
curvature of the one-forms $B_1^a$;   
$F_3^a$  and $F_1^a$  are the curvature of the two-forms $B_2^a$ and zero-forms
$S^a$ respectively.   The geometric nature of the indices $a,b$
is an interesting question. They can be  identified as $SO(4)$
indices:  one can assemble the six one-forms $B_1^a$ as components of a
Yang--Mills field valued in the Lie algebra of 
$SO(4)$ and    $B_2^b$ and $S^b$  as the components of a vector
of $SO(4)$. Thus, one  can   define  $F^a_{ij}$ as the non-Abelian curvature of 
$A$, and $F_1$ and $F_3$ as the covariant derivatives of the zero-forms $S^a$  and  two-forms   $B_2^a$. Having a topological theory with $SO(4)$ Yang--Mills invariance suggests a link to 
four dimensional topological gravity.

 We are now ready to construct  a five-dimensional theory associated to that of a four dimensional
theory for    
$B_{ij}$,   $\bar B_{ij}$,  $S
$ and $\bar S$, as an illustration of the algorithm relating $D$ and $(D+1)$ theories.

Because we have introduced the scalars $S$ and $\bar S$, we must complement
\symQf\  for
$D=4$ by
\def\S{{S}}
\eqn\titi{\eqalign{& Q \S=\Psi_\S
 \quad \quad \quad \quad 
Q \Psi_\S=0
}}
and  \symQdd\   for $D=5$ by 
\eqn\titii{\eqalign{&
Q \S=\Psi_\S
 \quad \quad \quad \quad 
Q \Psi_\S=\pa_5 \S
}}
with analogous  equations for $\bar S$. 

The four-dimensional theory relies
on the "mixed"  self-duality equations:
\eqn\selfB{\eqalign{&
\pa _{i } S= \e _{ijkl}\pa _{j }\bar B_{ kl}
  }} 
\eqn\selfBB{\eqalign{&
\pa _{i } \bar S= \e _{ijkl}\pa _{j } B_{ kl}
  }}
The classical topological action is 
\eqn\actionfour{\eqalign{&
\I_{4 \ \rm{top} }  = \int_4 \ dS\wedge d\bar B_2
+
d\bar S\wedge d  B_2
  }}
Using the gauge functions  
\selfB\ and \selfBB\ in the standard way (there are, actually, four
antighosts $\kappa^{-1}_a$ and  four
Lagrange multipliers  $\H   _a$ for the system $(S,B_2)$, which correspond to
the 1 degree   of freedom for   $S$ plus  the $ 3=6-4+1$ off-shell 
   gauge-invariant  degrees of  freedom  for the  two-form
$B_2$ in four
dimensions), 
one gets a four-dimensional TQFT
action \eqn\actionff{\eqalign{&
\I_{4     }  = \int_4 d^4x  (|\pa_i S |^2+
|\pa_i \bar  S |^2
+
|\pa_{[i} \bar  B_{jk]} |^2
+
|\pa_{[i}    B_{jk]} |^2
+\ldots)
}}
The
terms made explicit in this action  are obtained by  squaring  
the gauge functions.  The coefficients are chosen such that  the   mixed terms obtained  in this operation cancel  against 
$\I_{4 \ \rm{top} }$, leaving the squared curvatures. The   terms $\ldots$ stand for the various ghost-depending terms analogous to those made
explicit in  \west\ for the six-dimensional TQFT of two-forms.   We expect
that the more complete  TQFT action  using all fields appearing in
\reduc\ for
$D=4$ can be untwisted in an ordinary supersymmetric action, since this is
presumably the case for the eight-dimensional action for a three-form
\west.

The five-dimensional action  
\actiondd\ has a CS term
\eqn\actioncsfive{\eqalign{&
\I_{5 \ \rm{CS} }  = \int_5 \ B_2 \wedge d\bar B_2
+d^5 x \e^{ijkl5}(
\bar \varphi^+_i \pa_{[j} B_{jk]} 
+ \varphi^+_i \pa_{[j} \bar B_{jk]}
+\bar \Psi_{ij}  \Psi_{kl}
)
  }}
It   depends on the four-dimensional two-forms lifted to  five dimensions, but not on the scalars $S$. The terms depending on $  \varphi$ and $\bar \varphi$ 
are related to   Gauss's law for the
CS-like term $\ B_2 \wedge d\bar B_2$.

The  expression of  the $Q$-exact term  in  \actiondd\  is inspired from the
 four-dimensional gauge functions
\selfB\ and \selfBB. It is
\eqn\actiongsfive{\eqalign{ 
\int d^5 x \ \big\{\ 
Q, &
\chi_i (\e^{ijkl5}
\pa_{j} \bar B_{ kl} 
+\pa_{i} \bar S 
+\demi
\H_i )
+
\bar\chi_i (\e^{ijkl5}
\pa_{j}   B_{kl} 
+\pa_{i}   S
+\demi
\bar \H_i )
\cr
&
+
\Psi^-_{\mu\nu}
(\pa_{[5}  \bar  B_{\m\n]}+
\pa_{[\mu}  \bar  \varphi^+_{ \n]}
+
\Psi^-_{S}
 \pa_{ 5} \bar  S)
+
\bar\Psi^-_{\mu\nu}
(\pa_{[5}      B_{\m\n]}+
\pa_{[\mu}  r  \varphi^+_{ \n]}
+
\bar \Psi^-_{S}
 \pa_{ 5}    S )\ \}
}}
Adding the two terms \actioncsfive\ and \actiongsfive, we obtain
 \eqn\acfive{\eqalign{ 
\I_{5   }  = \int_5
 &
 B_2 \wedge d\bar B_2
+d^5 x \e^{ijkl5}(
\bar \varphi^+_i \pa_{[j} B_{kl]} 
+ \varphi^+_i \pa_{[j} \bar B_{kl]}
+\bar \Psi_{ij}  \Psi_{kl}
)
\cr
&
+  d^5x ( |\pa_\mu S |^2+
|\pa_\mu \bar  S |^2
+
|\pa_{[\mu} \bar  B_{\nu\rho]} |^2
+
|\pa_{[\mu}    B_{\nu\rho]} |^2
+\ldots)
 }}
The ... terms stand for terms depending on the topological ghosts, which are
 lengthy to write.
 Their presence  enforces the topological behaviour of the theory because
there are  
compensations between bosons and fermions. What truly counts  when one computes
$Q$-invariant observables are   the contributions  of the  zero modes that 
encode the topological information of the theory.  In the absence of the $Q$-exact term 
\actiongsfive\   it seems difficult to directly quantize the action \actioncsfive\ alone, and
the compensation between bosons and fermions is not explicit. In this sense, the action
\actiongsfive\ is a useful topogical gauge-fixing term for the theory.  

We could have considered a gauge in which the $Q$-exact terms are such that, after using the
constraints coming from field equations, only the term
$\int_5
 B_2 \wedge d\bar B_2
$  remains, while all fermions are spectators. We believe  that the definition of the system
from this sole term  could be  ambiguous  while  the complete action 
 \acfive\  determines a predictive QFT.

\subsec{Generalization}

We  generalize    to the case of a
$p$-form $B_p$ with curvature $F_{p+1}$. We denote as $X_{\do,r}$ the
$X_{\do,\m_1\ldots \m_{r}}$ component  of a $(r+1)$-form $X_{r+1}$. Then, the
following definition  generalizes  the expression  \symQdd\ of $Q$ that we have 
established for the case of a two-form
\eqn\symQdddd{\eqalign{& Q B_{p} =
 \Psi^-_{p}
\cr & 
 Q \Psi^-_{p} = F_{ \do,p} + d\phi^+_{p-1}
\cr & 
Q \phi^+_{p-1}
= \Psi^-_{\do,p-1} + d\phi^-_{p-2}
\cr & 
Q \phi^-_{p-2}
= \phi^+_{\do,p-2} + d\phi^-_{p-3}
\cr &
\ldots
\cr &
Q \phi^\pm_{O}
= \phi^\mp_{\do }  
}} 
We could as well consider a collection of $p$-forms, with different values of $p$. By construction, the operator $Q$ defined in \symQdddd\ satisfies the
relation \nildd. One can then obtain the $Q$-exact term of the  $(D+1)$-dimensional theory if one has      self
duality equations   for the $p$-form in $D$ dimensions. One must also investigate whether one has $Q$-invariant and  not $Q$-exact terms depending on the $p$-forms to eventually  get a
(supersymmetric) Chern--Simons action generalizing
\actioncsfive.

As new examples, we can start from      pairs of {  independent}
forms $B_p$ and   $C_{D-2-p}$, with different values of $p$. Indeed,  one may consider
combinations of the following   actions in $D$ and $(D+1)$ dimension
\eqn\acd{\eqalign{& \I_{D}
=\int_ {D} dB_p\wedge dC_{D-2-p} +\big\{Q,\ \}  
}} 
with their possible $(D+1)$ counterparts 
\eqn\acdd{\eqalign{& \I_{D+1}
=\int_ {D+1} 
\I_{D+1 \rm{CS}}[B,C,\Psi] +\big\{Q,\ \} } 
}
The expressions for $Q$ satisfy  \nild\ in \acd\ and  \nildd\
in \acdd. The expression of $\I_{D+1 \rm{CS}}[B,C,\Psi]$ depends on the particular cases and, to our knowledge, cannot be inferred from the existence of invariants in $D$ dimensions. 

There is another   
  $(D-1)$-dimensional TQFT action, which is :
 \eqn\acddd{\eqalign{& \I_{D-1} =\int_ {D-1}
B_p\wedge dC_{D-2-p} +\big\{Q,\ \}   }}
Its study would imply the analysis of the TQFTs in $D-2$ dimensions.

The action \acd\ induces a  TQFT that parallels the four-dimensional Yang--Mills case if one uses uses    as a gauge function the   duality condition between 
the curvatures  $ F_{ p+1}$ and $G_{d -1-p}$ of the  $p$-form gauge field
$B_p$ and   $B_{d -2-p}$:
\eqn\tilt{\eqalign{&  F_{ p+1} =^*G_{d -1-p} 
}}  
The four-dimensional Yang--Mills case is exceptional, since the self-duality
equation only involves one field (see \laroche\ for an attempt at classifying
the possible TQFT  with a single field). 

 Having  at our disposal   theories defined   by   \acd\   is a
new feature.  They  depend
on two independent fields which can be related by a self-duality relation between  their curvatures.   One can write all sorts of descent equations, which formally 
determine  observables and it would be    interesting   to investigate in which cases  these
TQFTs can be untwisted into supersymmetric theories.

Up to renamings, the field degrees of freedom
are the same for the actions \acd\ and  \acdd.  Let us   
conclude this section  by the following remark. The ordinary ghosts and
topological ghosts of $B_p$ and   $C_{D-2-p}$ can be unified in  the
expansion in ghost number of generalized forms $\tilde B_p$ and   $\tilde  C_{D-2-p}$
and   $\tilde  F_{p+1}$ and $\tilde  G_{D-1-p}$.  For these forms, the grading is   the sum
of $D$-dimensional  Lorentz degree and ghost number. If,     as explained in \kyoto, we accept fields with
negative ghost number in the generalized forms, one   finds that the  components  with
negative ghost number   in $\tilde  G_{D-1-p}$ are the Batalin--Vilkovisky    antifields of the fields  with positive ghost
number in   $\tilde B_p$, while 
the  components  with
negative ghost number   in $\tilde B_p$ are the Batalin--Vilkovisky    antifields of the fields  with positive ghost
number in   $\tilde  G_{D-1-p}$.  There are analogous   relations
between the field contents of $\tilde  C_{D-2-p}$ and  $\tilde  F_{p+1}$.

\newsec{Applications and gauging of free self-duality equations by dimensional
reduction}

As an example for the action \acd,  we have in $D=11$ the following classical 
"topological" term,   inspired by the $N=1,D=11$ supergravity:
 \eqn\sugra{\eqalign{& \I_{11} =\int_ {11} dC_3\wedge dB_6 + \big\{Q,\
\}     }}  and the gauge function in eleven dimensions
 \eqn\sug{\eqalign{&  dC_3 =^*dB_6   }}
Interactions can be introduced, e.g. by improving the gauge functions by addition
of a gravitational or Yang--Mills  Chern class of rank four \bks. The possibility of this modification
relies on the fact that  $C_3$ can be defined as  a connection rather than as a
form, as explained in \simons.  Thus we expect a TQFT in  $D=11$ and $D=12$, depending on a three-form and a
six-form $B_6 $, by the method  of section 2.

Let us now consider   the generic case for \acd, with 
the   free self-duality equation
\eqn\sugpp{\eqalign{&  dB_p =^*dC_{D-2-p}   }}
and  see in more detail the possibility of
dimensional reduction  briefly mentioned in section 2.1 for  the TQFTs 
determined  from  $\int_D  dB_p\wedge dC_{D-2-p}$. The foregoing analysis includes the particular case $D=2p+2$, in which a single one-form  $B=C$ is sufficient to write the self-duality equation.

 Dimensional  reduction  increases
the number of forms of a given degree  according to the property that a form of
degree $p$ in
$d$ dimensions gives in
 ($d-1$) dimensions a     form of degree $p$  and a     form of degree
$p-1$. The coefficients in the  following  Pascal triangle describe the number of forms
$B_{p}$, 
$B_{p-1}$, etc, obtained by dimensional reduction  of a $p$-form $B_{p}$ from
$ D$
to any given lower dimension $d=D-\Delta D$\foot{
The entries in the table vanish for forms that do not exist in the chosen reduced dimension.}:
\eqn\reducf{\eqalign{\matrix{     &   
{   B_{  p }} &
{   B_{  p-1 }} &
{   B_{  p-2 }} &
{   B_{  p-3 }} &
{   B_{  p-4 }} &
{   B_{  p-6 }} &
{   B_{  p-7 }} &
\ldots
\cr   
    D &      1
 \cr
   D-1    &   1&    1 \cr
      D-2   &   1&  2 &  1 \cr
      D-3&     1&  3 & 3  &  1 \cr
    D-4&     1&  4 & 6 & 4& 1 \cr
     D-5 &    1&  5 & 10&   10& 5 &  1 \cr
   D-6   &   1&  6 & 15& 20 & 15&6 &  1 \cr
   D-7    &   1&  7 & 21&   35& 35&21 &  7 &      1 \cr
   D-8     &   1&  8 & 28&   56& 70 &56 & 
28 & 8 &  1
\cr
   D-9    &   1&  9 & 36&   84& 126&126 &  84   & 36&  9 &  1
\cr
   D-10    &   1&  10 & 45&   120& 210&      252  
&  210 &    120 &   45 &   10 &  1 \cr
   D-11    &   1&  11 & 55&   165& 330&462 &  462 &    330 &   165 &  
 55 &   11 &  1 \cr
   D-12    &   1&  12 & 66&   220& 495&792 &     924
 &    792 &   495 &  
 220 &     66  &   12 &  1 \cr \dots &\cr\cr}
 }}
Clearly, the general formula for the number of $(p-\Delta p)$-forms obtained by
dimensional  reduction of a    $p$-form from $D$  to $d=D-\Delta D$ dimensions is 
$ 
C^{\Delta D}_{\Delta p}
$.

 Table \reducf\ indicates that, by    dimensionally reducing   from $D$ to $d $   a pair of
independent forms 
$B_p$ and $C_{D-2-p}$, one gets          $2 C^{\Delta D}_{\Delta p}$    forms of degree $p-\Delta
p$ and    $2 C^{\Delta D}_{\Delta p}$  forms of degree $d-2-(p-\Delta p)$. 
For  $D\neq 2p+2$, $B_p$ and 
$C_{D-2-p}$ contribute in an asymmetrical way  to the total number of forms of a given degree
in dimension $d$.  The  matching between the number of $q$-forms  and $(d-2-q)$-forms generated by
dimensional reduction  of $B_p$ and $C_{D-2-p}$  is important: it leads to multiple  self-
duality equations  in $d$ dimensions 
\eqn\comboo{\eqalign{dB^i_q=^*dC^i_{d-2-q}}}
where the index $i$ runs over the number of possibilities indicated by the combinatorial factor $ 
C^{\Delta D}_{\Delta p}
$.

 We will  use the fact that     the $d$-dimensional forms $B^i_q$  and $C^i_q$ can be
 arranged in a $SO(d)$-covariant way   as elements of representations of a Lie algebra
${\it G  }$. There are generally several possibilities for ${\it G  }$.
If the number of one-forms  $B_1^a$ and $C_1^a$ equals  the  dimension  of the  adjoint representation of
${\it {G } }$,  we can identify   $B_1^a$ and $C_1^a$ as the components of a 
${\it {G } }$-valued Yang--Mills field $A_1$,
and   gauge the global symmetry. This  supposes that   {\it {all} } other forms  $B_q^i$ and   $C_{d-2-q}^i$ fit in  
 various   representations of  ${\it {G } }$.  If it is the case,  
denoting   by $f^a_{bc}$ the structure coefficients   of  ${\it G  }$
and   by  $T^i_{ja}$   the matrix elements of  a given representation $X^i$ of 
${\it G   }$, the gauging simply      amounts 
to changing   the free dimensionally reduced   self-duality equations $
dB_q^i =^*dC^i_ {d-q-2}$   into the non-Abelian equations 
\eqn\drnaa{\eqalign{& 
 dB_q^i +T^i_{ja} A_1^a B_q^j =^*(dC^i_ {d-q-2}+ T^i_{ja} A_1^a  C_ {d-q-2}^j)
\cr &
  dA_1^a+{1\over 2} f^a_{bc} A_1^b A_1^c =^*(dC^a_{d-3} +    f^a_{bc} A_1^b   C^c_{d-3}      
)}} 
These equations are the self-duality equations  \tarata\ quoted in the  introduction.

 Imposing the  non-Abelian  self-duality equations \drnaa\
 as topological gauge conditions of a TQFT with a   $Q$-invariant action  is  
straightforward from the point of view of BRST quantization. The result is an   action 
\eqn\drna{\eqalign{& 
 \int d^d x\  \tr (|F^a|^2 +   \sum _q |DB^i_ {d-q-2} |^2+ |  D B_q^i|^2 +{\rm supersymmetric \ terms} )      
 }} 
The gauge function  involving the  
Yang--Mills  curvature  $F =dA+AA$ in \drnaa\ is a generalization of the three-dimensional   Bogomolny equation between a Yang--Mills field and a Lie algebra
valued scalar field. When the dimensional reduction is done  in four dimensions,
the gauge function  on   $F $ reduces to the self-duality condition 
$F=^*F$.

In the process that we have just described, the interactions only arise through minimal coupling to a Yang--Mills field.   However, if we were
to modify the free self-duality equations  before dimensional reduction, for instance by
adding to the free curvatures     Chern--Simons-like terms made of external products of various fields, the
gauging after dimensional reduction would  still be consistent. One  example  is to begin with a ten-dimensional self-duality equation   for a  four-form $B_4$ interacting with a pair of real  two-forms $B_2$ and $B'_2$,  
$G_5=^*G_5$, with 
$G_5=dB_4+  dB_2\wedge B_2'-dB'_2\wedge B_2$. The curvatures of both two-forms are  purely Abelian, $G_3=dB_2$   and $G'_3=dB'_2$.  By dimensional reduction in four
dimensions we obtain a TQFT whose gauge functions are  self-duality equations of the type  \selfB\
and \selfBB. The interacting terms  can be
worked out from the definition of $G_5$.  Another example is to begin with self-duality
equations involving $p$-forms with  Chapline--Manton-type  curvatures  
$G_{p+1}=dB_p+Q_{p+1}(A,F)$, where $Q_{p+1}(A,F)$  is  a Yang--Mills Chern--Simons
term. We expect a dimensionally reduced    theory  with  a gauge symmetry equal to  the
product of the original Yang--Mills symmetry by the  
symmetries coming from   dimensional reduction.  

 Note that a direct construction can yield  more invariants  in $d$ dimensions than those
provided by    dimensional   reduction of  the  invariants in $D$ dimensions. For instance, if
we start in $D=8$ with a  three-form $B_3$,  dimensional reduction  gives  four  one-forms
$B_1^a$ in $d=4$;      the    invariant  $\int _8 dB_3\wedge d
B_3$ gives 
$\int _4  \epsilon _{abcd} F_2^{ab}\wedge F_2^{cd}$,          where the internal indices
$a,b,...$ are $SO(4)$ indices.  However,  the  invariant  
$\int _4    F_2^{ab}\wedge F_ {2 ab}$ is not obtained by dimensional reduction, and must
be introduced directly in the four-dimensional theory.

Let us also note that the   global symmetry $SO(\Delta D)$, which generically occurs in
dimensional reduction, 
 is not the one that is  generally    used  for the
gauging.  There is however,  the  exceptional case  of the dimensional reduction of a 
three-form, which gives 
$\Delta D(\Delta  D-1)/2$ one-forms, which matches the dimension of the  adjoint
representation of  $SO(\Delta D)$.

As examples, let us first examine   the  five- and
four-dimensionsal gauge symmetries that are obtained by dimensional reductions of
$k$-forms   with a self-duality  equation $dB_k=*dB_k$ in  
$D=2k+2 $ dimensions. Using  \reducf,   we have,  for
$d=5$:
\eqn\reducfi{\eqalign{
D=6 &:  \ B_2 \to    
   d=5\ :   \ B_3  ,  \  B_2     
     \cr
D=8  &:  \ B_3 \to    
   d=5\ :   \ B_3  ,  \ 3B_2  ,   \  {\bf  3 B_1 }, \  S
     \cr   
 D=10  &:  \ B_4  \to    
   d=5\ :   \ B_4  ,  \ 5B_3   ,  \ 10B_2,  \  {\bf  10B_1 }, \  5S
     \cr   
 D=12 & :  \ B_5
   \to    
   d=5\ :  \ B_5, \ 7B_4   , \ 21B_3 ,    \ 35B_2,  \  {\bf 35 B_1 }, \  21S
\cr
 D=14&  :   \ B_6
   \to    
   d=5\ :    \ 9 B_5 ,   \ 36B_4   ,  \ 84 B_3,  \ 126 B_2 , \ 
{\bf126 B_1} ,  \ 84 S
     }
}
and for   $d=4$:
\eqn\reducfo{\eqalign{ 
D=6 &:  \ B_2 \to    
   d=4\ :   \ B_2   ,  ,   \  {\bf  2 B_1 },   \   S 
     \cr
D=8  &:  \ B_3 \to    
   d=4\ :   \ B_3   , \ 4B_2  ,   \  {\bf   6 B_1 }, \  4S
     \cr  
 D=10 &: \ B_4  \to    
   d=4\ :   \ B_4 ,   \ 6B_3 ,    \ 15B_2 , \  {\bf   20B_1 }, \  15S
   \cr
 D=12 &: \ B_5
   \to    
   d=4\ :   \ 8B_4   , \ 28B_3 ,    \ 56B_2 , \  {\bf  70 B_1 }, \  56S
\cr
D=14&  : \ B_6   
   \to   
   d=4\  :    \ 45 B_4  ,   \ 120 B_3,  \ 210 B_2, \ 
{\bf 252 B_1} ,\ 210 S
     }}
We have put in bold
characters the number of one-forms that occur for each
decomposition.  Thus,   possible internal symmetries such that the number of the one-
forms fits the number of generators of their Lie algebra are,    in   $d=5$:
 \eqn\liealgfi{\eqalign{ 
3 B_1 &: SU(2)   \cr
10 B_1 &: SO(5); \   SO(4)\times SU(2) \times U(1)      \cr
35  B_1 &: SU(6); \   SO(7)\times G_2    \cr
126 B_1 &: E_6\times SU(7) ;  \ SU(8)  \times SU(8)    \cr
     }}
and in   $d=4$:
 \eqn\liealgfo{\eqalign{ 
6 B_1 &: SO(4)   \cr
20 B_1 &: SO(5)\times SO(5); \   SO(4)\times  G_2     \cr
70  B_1 &: SU(6)\times SU(6) \  ;SU(2) \times U(1) \times SO(12); \  
;  \ SO(8)  \times SO(8) \times  G_2  \cr 252 B_1 &: E_8\times SU(2) \times U(1)    ; \ (E_6\times SU(7))^2  \cr
     }}
One can verify in each case that the other fields        can be arranged into     representations
of the Lie algebra ${\it G}$ as "matter" fields. Let us work out a few examples. For   $d=5$
with
${\it G}=SO(5)$,  the 10 one-forms coming from the dimensional reduction of
 a four-form in ten
dimensions  are identified as a
 Yang--Mills field for
$SO(5)$ while, the two-form are also valued in the Lie algebra of
$SO(5)$; the 5 three-forms and zero-forms $B_3$ and $S$ are in  the
fundamental representations and $B_4$ is a singlet.  For $d=5$ and  
${\it G}=  SO(7)\times G_2  $, 
    the $35=21+14$ one-forms   
coming from the dimensional reduction of a five-form in 12 dimensions build  a
Yang--Mills field valued in the adjoint representations of ${\it
G}=SO(7)\times G_2$;   the 35     two-forms can be arranged in the same
representation;  
 the 21 "matter" three-forms and zero-forms $B_3$ and $S$   can be identified   
either as  $\underline 21$,
 $\underline 7 \oplus \underline 7 \oplus  \underline 7 $ or 
$\underline 7 \oplus \underline 14$
  of $SO(7)\times G_2$; the seven four-forms $B_4$ are in  the fundamental
representation of $SO(7) $ or $  G_2$ and  $B_5$  is a singlet.  This five-dimensional case is
  of    interest, since  it contains a five-form giving an action $\int_5 B_5$, 
which can be used as a
Wess--Zumino action to compensate for a possible five-dimensional anomaly, and   
seven four-forms
$B_4$, which can be used to give  a scale \townsend.

Let us consider   theories with pairs of   fields, adapted to a starting point in twelve
dimensions, which is tantalizing from the point of view of $F$ theory. For instance,  if we
start from a six-form and a three-form in twelve  dimensions, in the context of  the CS
supersymmetric  theory associated to  an eleven-dimensional TQFT stemming from
\sugra. Table 
\reducf\ gives the following field spectrum in four dimensions
\eqn\reduconze{\eqalign{    
D=12&  : \ B_6   
   \to   
  d=4\  :    \ 28 B_4  ,   \ 56 B_3,  \ 70 B_2, \ 
{\bf 56 B_1} ,\ 28 S
\cr
 D=12&  : \ B_3   
   \to   
  d=4\  :        \ 1B_3,  \ 28 B_2, \ 
{\bf 28 B_1} ,\ 56 S
     }}
We see that one has the possibility of having a four-dimensional
theory with the one-forms in  the adjoint representations $\underline
 {28}\oplus
\underline { 28}  \oplus
\underline { 28}$ of
$SO(8)\times SO(8)\times SO(8)$.

Another possibility is to start  from a TQFT with  a  six-form
and a four-form in twelve  dimensions, with the  action 
$\int_{12} dB_6 \wedge dB_4$. Then
\eqn\reduconze{\eqalign{    
D=12&  : \ B_6   
   \to   
   d=4\  :    \ 28 B_4  ,   \ 56 B_3,  \ 70 B_2, \ 
{\bf 56 B_1} ,\ 28 S
\cr
D=12&  :  \ B_4   
   \to   
  d=4\  :        \ 1B_4,  \ 28 B_3, \ 28 B_2, \
{\bf 56 B_1} ,\ 70 S
     }}
In this case, the gauge symmetry can be chosen for example as  $[SO(8)]^4
$ for the four-
dimensional theory.

 Finally,  we can consider   a theory with a seven-form
and a three-form TQFT in twelve  dimensions,  with the  action $\int _{12} (dB_7 \wedge dB_3 + dB_3  \wedge dB_3
\wedge dB_3)$. Then, 
\eqn\reduconze{\eqalign{    
D=12&  : \ B_7   
   \to   
  d=4\  :       \ 56 B_4,  \ 70 B_3, \  56 B_2  , 
{\bf 28 B_1} ,\  8 S
\cr
D=12&  :  \ B_3   
   \to   
  d=4\  :        \ 1B_3,  \ 28 B_2, \ 28 B_1, \
{\bf 28 B_1} ,\ 56 S
     }}
The gauge symmetry can be chosen for example  as  $SO(8)\times SO(8)$ or
$ SO(3)^6\times SO(10)^2$
for the four-dimensional theory.

 We conclude this section by the following remarks. The  gauging  of  global
internal symmetries in TQFTs when we dimensionally reduce the free self-duality
equation shown above is analogous to   that occurring   in supergravities. For 
supergravities,        string inspired  arguments justify this gauging \sugragauging.  

 Untwisting
the TQFTs obtained by dimensional reduction of free theories as those given by
\reducfi\ and
\reducfo\ can give theories with ordinary Poincar\'e supersymmetry, by
redefining the topological ghosts   as the elements of classified
multiplets of supersymmetry. There are examples of this in \west. In this case,
the free
$Q$-invariant actions that use the Abelian  self-duality equations as
topological gauge functions     satisfy Poincar\'e supersymmetry
after   untwisting the fields. The generator of the
$Q$-symmetry appears as a particular combination of the generators of
supersymmetry, i.e. the parameter of the BRST symmetry is a one-dimensional projection of the spinorial parameter  of Poincar\'e supersymmetry. 
The
$Q$-symmetry survives, by construction, the untwisting procedure and the
introduction of interactions. However, when one enforces
 in a
   $Q$-invariant way
the gauging of internal symmetries by    modification of the gauge functions,  
it is     unclear to us if   the 
  the {\it full} supersymmetry of the action is still existing after untwisting
the fields as in the free case.  If a reduction of supersymmetry  occurs, it could
parallel  the analogous property at work
when one tries to
find an effective supersymmetric theory in the  world volumes  
of branes. In our context, the extreme case would be that only the $Q$ symmetry is
present in the interactive theories. 
\vskip 0.5cm

\noindent{\bf Acknowledgements}
We wish to thank S. Ferrara, C. Kounnas  and S. Shatashvili for discussions.

 \listrefs 
\bye